# A Distance Metric Learning Model Based On Variational Information Bottleneck

YaoDan Zhang [1], Zidong Wang[2], Ru Jia[3(✉)] and Ru Li[4]

[1] Beijing University of Posts and Telecommunications, Beijing, China
zyd_cs@bupt.edu.cn
[2] Central South University, Changsha, China
studentwangzidong@163.com
[3] Inner Mongolia University, Hohhot, China
csjiaru@imu.edu.cn
[4] Inner Mongolia University, Hohhot, China
csliru@imu.edu.cn

**Abstract.** In recent years, personalized recommendation technology has flourished and become one of the hot research directions. The matrix factorization model and the metric learning model which proposed successively have been widely studied and applied. The latter uses the Euclidean distance instead of the dot product used by the former to measure the latent space vector. While avoiding the shortcomings of the dot product, the assumption of Euclidean distance is neglected, resulting in limited recommendation quality of the model. In order to solve this problem, this paper combines the Variationl Information Bottleneck with metric learning model for the first time, and proposes a new metric learning model VIB-DML (Variational Information Bottleneck Distance Metric Learning) for rating prediction, which limits the mutual information of the latent space feature vector to improve the robustness of the model and satisfy the assumption of Euclidean distance by decoupling the latent space feature vector. In this paper, the experimental results are compared with the root mean square error (RMSE) on the three public datasets. The results show that the generalization ability of VIB-DML is excellent. Compared with the general metric learning model MetricF, the prediction error is reduced by 7.29%. Finally, the paper proves the strong robustness of VIB-DML through experiments.

## 1 Introduction

In the past ten years, the amount of data on the Internet has been increasing everytime and it is difficult for users to obtain effective information from the mass data. For companies, how to target products and services to users has become a huge challenge today. Recommender systems can meet the individual needs of both sides, and has been widely studied and used.

Collaborative filtering algorithm is one of the algorithms widely used in recommendation systems in recent years [1]. The matrix factorization model (MF) in the collaborative filtering algorithm has been extensively studied, and many variants have been derived [2]. Such as Double Topics with Matrix Factorization (DTMF) [3], Bias Singular Value Decomposition (BiasSVD) [1], Probabilistic Matrix Factorization (PMF) [4], Bayesian Probabilistic Matrix Factorization [5], Trust-Based Matrix Factorization (Trust-MF) [6] and so on. The above models all use dot products to calculate the relationship between users and items [1,3-6]. In the recommendation scence, using dot products which do not satisfy the triangle inequality do not satisfy the definition of the distance function, and cannot accurately measure the interaction relationship between the user and the item, resulting in the user vector and item vector optimize towards the wrong direction. Therefore resulting invector suboptimization and the increasing of errors on the model [7,8].To sovle the problem, Hsiehet et al. proposed Collaborative Metric Learning (CML) [9]. CML uses the Euclidean distance that satisfies the triangle inequality instead of the dot product to measure the relationship between the user vector and the item vector. And this method increases distance between users and items that they do not like and corrects the vector calculation errors caused by using the dot product.However, CML cannot narrow the distance between users and their favorite items,which may cause the favorite items to be too far away from the user, and the model makes incorrect recommendations [7]. Metric Factorization (MetricF) proposed by Zhang et al. [7] is improved on the basis of CML, and narrows the distance between users and their favorite items by the distance space decomposition. The above models which have all used the Euclidean distance to calculate the relationship between users and items are collectively called the Distance Metric Learning (DML) and as the current mainstream recommendation models, they are applied to the field of rating prediction [7], Top-N recommendation [9], sequential recommendation [10], context recommendation [11]. DML solves the vector suboptimization problem caused by the use of dot products in matrix factorization models, but it also introduces new problems: 1) DML uses the embedding matrix to generate latent vectors, which causes the latent vectors in space to be too sensitive to the dimensionality of the feature space. And there may be some redundant information in the latent vectors,which making the model overfitting [12]. 2) DML does not consider that the data needs to satisfy the characteristics of independently identically distribution, which is the assumption that Euclidean distance depends on [13]. Failure to satisfy the above assumptions will lead to errors in the calculation of feature vectors, and the generalization ability of the model may be damaged [14]. Principal Components Analysis (PCA) is a kind of dimensionality reduction algorithms, which maps high-dimensional features to lower dimensions, removes the redundancy of feature vectors while ensuring the independently identically distribution of data after dimensionality reduction [15]. However, in the recommendation scence that requires data label information, the model cannot be trained normally because of the fact that PCA is unsupervised makes itself unable to blend into DML which is supervised [16]. In contrast, Variational Information Bottleneck (VIB) [12] can be blended into DML while solving the problems of DML, and the entire model can be trained normally [17]. VIB can extract important feature vectors and align the distribution of the latent variables with the normal distribution by measuring the posterior distribution of the latent variables and the KL divergence of the standard normal distribution [18], so that the latent variables remain independent and identically distributed. VIB uses mutual information to measure the amount of information transmitted by the model during training, and use variational inference to derive a variational lower bound of mutual information. The characteristics of VIB are as follows: 1) The latent vectors of VIB are bound by mutual information. This method removes redundant information in the latent vectors. 2) Decouple the latent vectors, and satisfy the assumption of Euclidean distance. VIB's own peculiarities can solve the two problems of DML. Therefore, this paper combines the theory of Variational Information Bottleneck, proposes a rating prediction algorithm called VIB-DML based on Variational Information Bottleneck. This algorithm eliminates redundancy of the feature vectors and improve the robustness of the model, use Kullbac-Leibler divergence [18] to constrain the feature vectors to obey standard and normal distribution and satisfy the assumption of Euclidean distance.

## 2    A Distance Metric Learning Model Based On Variational Information Bottleneck

Aiming at the characteristics of Euclidean distance that requires data to be independent and identically distributed and the requirement for eliminating redundancy of information, this chapter presents a distance metric learning model based on variational information bottleneck called VIB-DML. Tab. 1 shows the symbol description of VIB-DML.

**Table 1.** Notations used in VIB-DML.

| Symbol | Description |
| --- | --- |
| $x, y, z$ | Input variables $x$, label values $y$, latent variables $z$. |
| $I$ | Mutual information. |
| $P_u$ | The latent feature vector of user $u$. |
| $Q_i$ | The latent feature vector of item $i$ |
| $R_{ui}$ | The real rating of user $u$ to item $i$. |
| $\widehat{R}_{ui}$ | The prediction rating of user $u$ to item $i$. |
| $R$ | The observed matrix. |
| $D_{ui}$ | The true distance between user $u$ and item $i$. |
| $\widehat{D}_{ui}$ | The prediction distance between user $u$ and item $i$. |
| $KL$ | Relative entropy, A.K.A. Kullback-Leibler divergence. |
| $\theta$ | Probability distribution parameter to be estimated. |
| $\beta$ | Hyperparameter,to controll the proportion of penalty term. |
| $q(z)$ | The prior distribution of latent variable $z$. |
| $b_u$ | The bias term of user $u$. |
| $b_i$ | The bias term of item $i$. |
| $R_{global}$ | Global average rating. |
| $k$ | The dimension of feature vector. |

## 2.1 Constraints on Feature Vectors

In the recommendation scence, the goal of this article is to estimate the unrated user-item matrix under the condition of the existing observation rating matrix, which iterate the most appropriate parameter $\theta$ for the observed matrix. Project the user $u$ and the item $i$ into the same $k$-dimensional latent space to obtain the user vector $P_u$ and the item vector $Q_i$. The $P_u$ and $Q_i$ here is the parameter $\theta$ to be estimated. In this article, VIB is used to process $P_u$ and $Q_i$, and the latent variable $z$ generated by the VIB encoder is used as the new $P_u$ and $Q_i$. VIB-DML solves the two problems of DML through the following operations. 1) Eliminate the redundancy of feature vectors $P_u$ and $Q_i$. 2) Convert $P_u$ and $Q_i$ into independently and identically distributed vectors.

For the first point, VIB-DML separately limits user vectors and item vectors to complete the task well, which makes the model only allow the most important information to pass through the "bottleneck" in the optimization process. The latent variables $z_u$, $z_i$ generated by the VIB-DML's two encoders contain the most important information of the user feature vector $P_u$ and the item feature vector $Q_i$.

For the second point, VIB-DML uses VIB to optimize $KL(p(z|x)\| q(z))$ to force $p(z|x)$ approach $q(z)$, $q(z)$ is a multivariate standard normal distribution. After the optimization, the posterior distribution $p(z|x)$ of the latent variables $z_u$, $z_i$ satisfies multivariate standard normal distribution $N(\mu,\sigma^2)$. Here, $\mu$ represents a $k$-dimensional vector with a mean of 0, and $\sigma^2$ represents a $k\times k$ square matrix with all its main diagonal entries are equal to 1. As shown in Eq. 1, $p(z) = \int p(z|x)\,p(x)\,\boldsymbol{dx} = N(0,1)$, $\int p(x)\,\boldsymbol{dx} = N(0,1)$, also obeys the multivariate standard normal distribution at this time. Therefore, $P_u$ and $Q_i$ generated by VIB-DML coder are independent and identically distributed, thus the restrictions of using Euclidean distance are lifted.

$$p(z|x) \sim N(\mu,\sigma^2)\ s.t. \tag{1a}$$

$$\mu = [0_1,0_2,0_3,\ldots,0_k]^{\mathrm{T}} \tag{1b}$$

$$\sigma^2 = [1_1,1_2,1_3,\ldots,1_k]^{\mathrm{T}} \tag{1c}$$

After obtaining the new user vector $P_u$ and the new item vector $Q_i$, Euclidean distance is used to measure the distance between the user $u$ and the item $i$, as shown in Eq. 2. Because it is necessary to compare the error between the predicted Euclidean distance and the existing rating label, the predicted Euclidean distance is converted into the ra-ting given by user $u$ to item $i$. The conversion of the rating follows the method of reference [6], which is defined as Eq. 3.

$$\widehat{D}_{ui} = (P_u - Q_i)^2 = \sqrt{\sum_{k=1}^{k} (P_{uk} - Q_{ik})^2} \qquad (2)$$

$$\widehat{R}_{ui} = R_{max} - \widehat{D}_{ui} \qquad (3)$$

## 2.2 Objective Function Optimization

VIB-DML encodes the latent variables $z_u$ of user $u$ and $z_i$ of items $i$ through two neural networks, namely $P_u$ and $Q_i$ in Eq. 2. The initial objective function is as follows:

$$arg \max_{\theta} I(z_u, z_i, y | \theta) - \beta_1 I(x_u, z_u | \theta) - \beta_2 I(x_i, z_i | \theta) \qquad (4)$$

For the first term of Eq. 4,expand $KL(\, p(y | \, z_i, \, z_i) \| q(y | \, z_i, \, z_i)) \geq 0$,to get variational lower bound of $I(z_u, z_i, y | \theta)$, as in Eq. 5.

$$\iiint p(z_u, z_i, y) \, log \frac{q(y | z_u, z_i)}{p(y)} \, dy \, dz_u \, dz_i = \iiint p(z_u, z_i, y) \, log \, q(y | z_u, z_i) \, dy \, dz_u \, dz_i + E(Y) \qquad (5)$$

Expand $I(x_u, z_u)$ in Eq. 4 to obtain Eq. 6.

$$\iint p(x, z_u) \log \frac{p(z_u | x)}{p(z_u)} \, dx \, dz_u \qquad (6)$$

Because the latent variable distribution $p(z_u) = \int p(x) p(z_u | x) \, dx$ ,and the posterior distribution $p(z_u | x)$ is difficult to calculate.Therefore,consider a simple distribution $q(z)$ to approximate the posterior distribution $p(z_u | x)$, and Eq. 6 is transformed into Eq. 7:

$$\int p(x) KL(p(z_u | x) \| q(z_u)) dx - KL(p(z_u) \| q(z_u)) < \int p(x) KL(p(z_u | x) \| q(z_u)) dx \qquad (7)$$

The result of minimizing the variational upper bound $\int p(x) KL(p(z_u | x) \| q(z_u)) dx$ is similar to minimizing $I(x_u, z_u | \theta)$ and $I(x_i, z_i | \theta)$ .

The derivation of the third term $I(x_i, z_i | \theta)$ in Eq. 4 is the same as the derivation of the second term in Eq. 4.

From the above derivation, we get the theoretical framework of VIB-DML as follows:

$$\iiint p(z_u, z_i | x) \, log \, q(y | z_u, z_i) \, dz_u \, dz_i \, dx - \beta_1 \int p(x) KL(p(z_u | x) \| q(z_u)) dx$$

$$- \beta_2 \int p(x) KL(p(z_i | x) \| q(z_i)) dx \qquad (8)$$

Aiming at the first problem of DML,VIB-DML uses $\beta_1 I(x_u, z_u | \theta)$ and $\beta_2 I(x_i, z_i | \theta)$ as the regularization terms of the loss function, so that the model can eliminate the redundancy of information in the latent vector and extract the most important features for the task, then the first problem of DML was solved. $\beta_1$, $\beta_2$ indicate the intensity of eliminating the redundancy.

In response to the second problem of DML,VIB-DML sets the prior distributions $q(z_u)$ and $q(z_i)$ in Eq. 8 as multivariate standard normal distributions, meanwhile, sets $p(z_i|x)$ and $p(z_u|x)$ as a normal distribution. When the optimization is completed, the posterior distribution $p(z|x)$ is a multivariate standard normal distribution, and all components are independent and identically distributed. It satisfies the assumption of Euclidean distance and solves the second problem of DML.

The rating prediction task is a regression task, and assume $q(y|z_u,z_i)$ as the mean squared error function.At the same time,considering that the ratings are also related to users or items itself,the bias items $b_u,b_i,R_{global}$ proposed by Koren et al. [10] are added. In summary, the loss function of VIB-DML is shown in Eq. 9.

$$\mathcal{L} = \sum_{u,i \in R} \left( R_{u,i} - \left( R_{max} - \sqrt[2]{\sum_{k=1}^{n}(P_{uk} - Q_{ik})^2} + b_u + b_i + R_{global} \right) \right)^2$$
$$+ \frac{1}{2}[\beta_1(-log\,\sigma_u{}^2 + \mu_u{}^2 + \sigma_u{}^2 - 1) + \beta_2(-log\,\sigma_i{}^2 + \mu_i{}^2 + \sigma_i{}^2 - 1)] \qquad (9)$$

## 2.3 Algorithm Summary

The optimization process of VIB-DML is shown in algorithm 1. In the end, Adagrad [19] is used to update parameters.

**Table 2.** Algorithm Procedure

---
**Algorithm 1**: parameters solution of VIB-DML
---
**Input**: rating matrix $R$, user/item vector $\{P_u,Q_i\}$, the dimension of feature vectors $k$,bias items$\{b_u,b_i,R_{global}\}$ ,hyper-parameters $\{\beta_1, \beta_2\}$, the maximum of ratings $R_{max}$

**Output**: feature vectors $P_u,Q_i$, bias items $\{b_u,b_i\}$

**while** $P_u,Q_i$ is not converged **do**

    initialize $P_u,Q_i$ with normal distribution.

    initialize $b_u,b_i$ as n-dimensional zero vectors.

  **if** $(u,i,r) \in R$ **then**

      $z_u \leftarrow f_{user\;embedding}(P_u)$

      $z_i \leftarrow f_{item\;embedding}(Q_i)$

      $\widehat{D}_{ui} = \sqrt{(z_u - z_i)^2} = \sqrt{\sum_{k=1}^{n}(Z_{uk} - Z_{ik})^2}$

      $\mathcal{L} = \sum_{u,i \in R}(R_{u,i} - (R_{max} - \widehat{D}_{ui} + b_u + b_i + R_{global}))^2$
        $+ \frac{1}{2}[\beta_1(-log\,\sigma_u{}^2 + \mu_u{}^2 + \sigma_u{}^2 - 1) + \beta_2(-log\,\sigma_i{}^2 + \mu_i{}^2 + \sigma_i{}^2 - 1)]$

      $P_u,Q_i,b_u,b_i \leftarrow f_{Adagrad}(P_u,Q_i,b_u,b_i,\mathcal{L})$

  **end if**

**end while**

# 3    Evaluation

This chapter compares VIB-DML and related recommendation models on real datasets from four aspects: performance, parameters, robustness and feature vectors.

## 3.1    Datasets and Performance Metrics

This paper uses the movie rating dataset Movielens 100K [20], the general rating dataset Epinions [21] and the movie rating dataset FlimTrust [22]. The statistical results of the three datasets are shown in Tab. 3.

**Table 3.** Statistics of datasets.

| Dataset | NO.of users | NO.of items | NO.of ratings | the range of ratings |
|---|---|---|---|---|
| MovieLens 100K | 943 | 1682 | 100,000 | 1-5 |
| Epinions | 40163 | 139738 | 664824 | 1-5 |
| FilmTrust | 1508 | 2071 | 35497 | 0.5-4 |

This paper adopts the Root Mean Squared Error (RMSE), which is commonly used in the field of recommender systems. Compared with the Mean Squared Error (MSE), it excludes the influence of dimensions. The specific formula is as follows:

$$\text{RMSE} = \sqrt{\frac{\sum_{i=1}^{n}(\hat{R}_{ui} - R_{ui})^2}{n}} \tag{10}$$

## 3.2    VIB-DML vs. other approaches

We compare with four state-of-the-art approaches.

1)BiasSVD [1]: Improve singular value decomposition (SVD) and add some inherent attributes of users and items, such as the user's rating schemes and the true quality of items, which can be learned from parameters.

2)PMF [4]: A probability model is introduced on the basis of matrix decomposition. Assuming that the observation noise, user attributes, and item attributes are all Gaussian distributed, the model parameters can be estimated through maximum likelihood estimation and then the ratings can be predicted.

3)NNMF [23]: Aiming at the shortcoming that the dot product can only establish linear relationship, a neural network which can catch the nonlinear relationships between users and items with nonlinear activation functions is introduced to replace the dot product.

4)MetricF [7]: Because the distance directly defined by dot product does not satisfy the triangle inequality, Euclidean distance is used to replace dot product to establish the distance relationships between users and items.

### 3.3 Evaluatioin and Analysis

**Performance**. We use Hold-Out to divide datasets randomly 5 times. The training set accounts for 90% and the test set accounts for 10%. The final result that used to evaluate the generalization capability of a model is the average of the RMSE on the test set of 5 times random partition.

In this paper, we use the deep learning framework Keras to complete the experiment. Part of the baseline model realizes by using an open-source toolkit DeepRec [24]. We focus on the limitations of Euclidean distance. In the course of the experiment, the part of confidence is removed to reduce irrelevant factors when reproducing MetricF. Some experimental results about the influence of deferent epochs on the selected models are shown in Fig. 1.

The dimension $k$ of the feature vector is shared by four models, and $k$ directly determines the expression ability of users and items. And the value of $k$ is used to distinguish the experimental results. A part of the experimental parameters are set as follows: epoch = 50, learning rate = 0.05, batch size = 256.

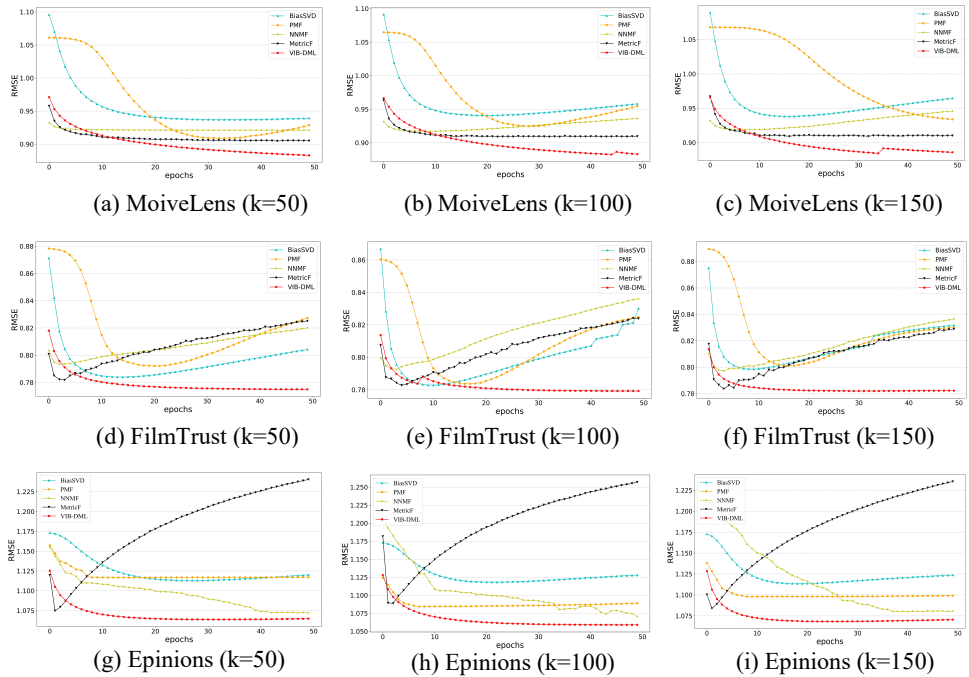

(a) MoiveLens (k=50)　　(b) MoiveLens (k=100)　　(c) MoiveLens (k=150)

(d) FilmTrust (k=50)　　(e) FilmTrust (k=100)　　(f) FilmTrust (k=150)

(g) Epinions (k=50)　　(h) Epinions (k=100)　　(i) Epinions (k=150)

**Fig. 1.** Some experimental results on selected datasets when k=50, k=100 and k=150.

It can be learned from Fig. 1 that NNMF converges faster, but it tends to overfit. PMF and BiasSVD are used as matrix factorization models, the error curve after convergence is comparatively smooth, but there is also a problem of over-fitting. MetricF and VIB-DML have excellent generalization capacity. VIB-DML converges faster and has high stability after convergence. Results from what have been discussed above can guide the selection of model parameters. After parameter optimization, in order to show the differences between the models more intuitively, the final experimental results are shown in Fig. 2.

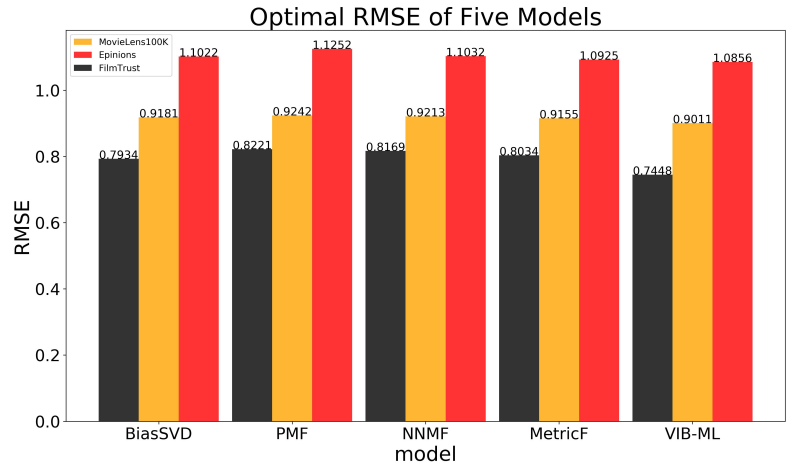

**Fig. 2.** Optimal RMSE of Five Models

As we can see in the Fig.2, PMF and BiasSVD, as matrix factorization models, are limited by the measure method of the dot product. Their averages of optimal RMSE on the three datasets are 4.36% and 2.25% higher than VIB-DML, respectively. As an improved nonlinear model of matrix factorization, NNMF uses neural networks instead of dot product for rating prediction, and its average optimal RMSE is 1.16% lower than PMF. This shows that neural networks may be superior to dot products in measuring the relationship between users and items. MetricF uses Euclidean distance to measure the relationship, and updates the position of users and items in space through distance metric learning. In that approach, linear correlation is found in dimensions and the RMSE has increased by 2.17% on average compared with VIB-DML. The model VIB-DML proposed in this paper has achieved the best results on the three real-world datasets. Compared with the main comparison model MetricF, the RMSE of VIB-DML is reduced by 7.29% at the maximum. By the experiment data we can see the generalization capability of VIB-DML is excellent. The above results show that the VIB-DML proposed in this paper can effectively measure the relationship between users and items on different datasets, and the recommendation results are better.

**Parameter Analysis**. *1)The dimension k of user vector and item vector.*

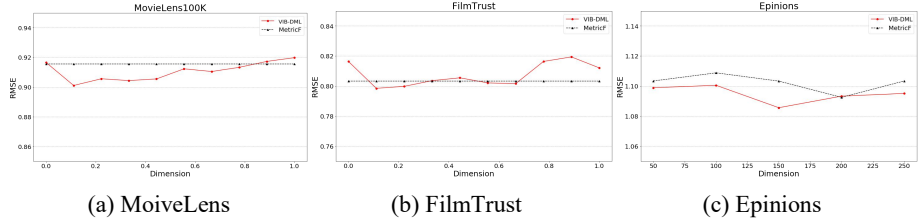

(a) MoiveLens      (b) FilmTrust      (c) Epinions

**Fig. 3.** Effect of dimension

As shown in Fig. 3, when running on different datasets (Movielens, FilmTrust, Epinions), the optimal RMSE of MetricF is obtained at k={150,100,200} while the optimal RMSE of VIB-DML is obtained at k={150,150,150}. In the same dataset, the optimal RMSE of VIB-DML and MetricF will both change when k changes, and the average RMSE of VIB-DML is lower than MetricF. The experimental results show that VIB-DML has more accuracy.

*2)The hyper-parameter $\beta$*

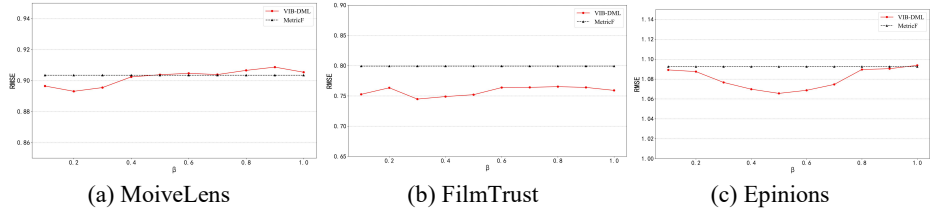

(a) MoiveLens      (b) FilmTrust      (c) Epinions

**Fig. 4.** Effect of hyper-parameter $\beta$

Fig. 4 shows the changes of hyper-parameter $\beta$ have effect on model performance. $\beta$ can control the amount of information flowing between $(x, z)$, and the larger the $\beta$, the less the amount of information flowing between $(x, z)$, thus playing the role of "information bottleneck". MetricF in the figure does not have hyper-parameter $\beta$, and its RMSE is set to the optimal RMSE in Fig. 2. For VIB-DML, the dimension running on the three datasets is set to the optimal dimension {150,150,150} in the section of parameter analysis. The three datasets have different sensitivity to $\beta$. In MovieLens 100k and FilmTrust, we obtain the minimum of the model error when $\beta = 0.2$ and $\beta = 0.3$, respectively. In Epinions, the model error reaches the minimum when $\beta = 0.5$. The above experiments show that $\beta$ is an important factor affecting model errors, and the difference of values in a single data set will lead to changes in model accuracy.

**Robustness**. VIB is characterized by eliminating the redundancy in feature vectors, and the generated feature vectors have better quality. Reference [12] pointed out that VIB has better robustness, and the value of hyper-parameter dimension k has little effect on it. The VIB-DML model combined with VIB should have good robustness in theory. This paper verifies this through the following experiments.

After weighing the models training time and the actual performance of the models in the section of performance, we choose BiasSVD and MetricF as comparisons, and also choose 500 as the dimension of feature vectors in the models. The experimental results are shown in Fig. 5.

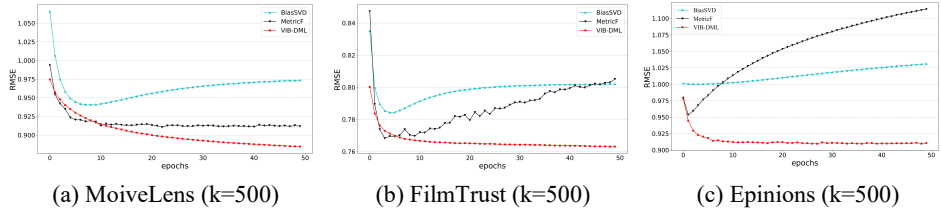

(a) MoiveLens (k=500)  (b) FilmTrust (k=500)  (c) Epinions (k=500)

**Fig. 5.** Robustness experiment

It can be seen from Fig. 1 and Fig. 5 that comparing the performance of the models under the condition of dimension k={150,100,200}, when k=500, the RMSE of the models BiasSVD, MetricF and VIB-DML on the three datasets increased by 5.54%, 4.21%, and 3.58% on average, respectively. According to reference [7,8], the change of RMSE indicates that the 500-dimensional feature vectors contains redundant information, because of that, both BiasSVD and MetricF can neither avoid the problem of over-fitting nor converge. VIB-DML can obtain the convergence when Epoch=20 on average, and the performance loss is only 3.58%. Experimental results show that VIB-DML can effectively eliminate the redundancy of feature vectors and has strong robustness, thus the over-fitting problem of DML models is solved.

**The Location of Vector in Space.** The DML in rating prediction scenario can link the distances between the vectors with the ratings. Only the approaches of DML models can be used for visual vector position experiments. Both MetricF and VIB-DML belong to the DML models. In this section, MetricF and VIB-DML are selected for contrast experiment.

Due to the limited optimization effect for MetricF, the relative locations of the generated vectors is incorrect. In this section, VIB-DML is used for comparison experiments. The results are shown in Fig. 6:

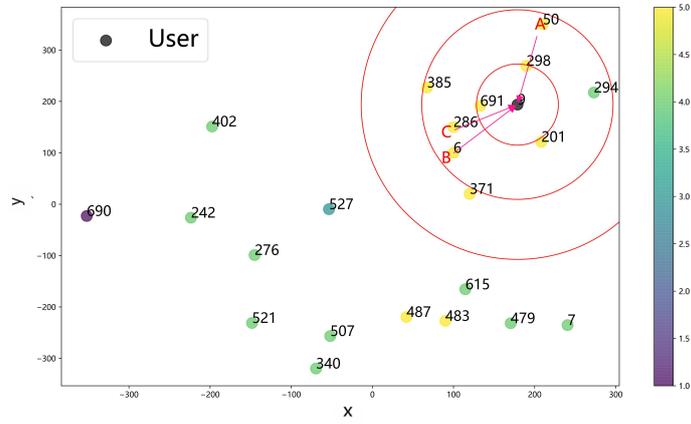

(a) VIB-DML

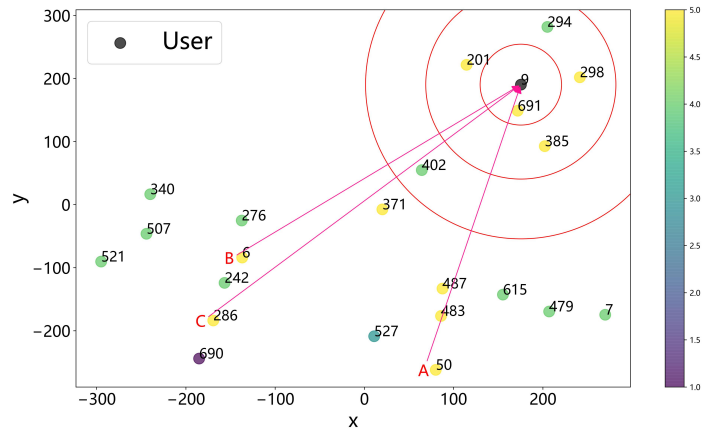

(b) MetricF

**Fig. 6.** The eigenvector distance relationship generated by VIB-DML (a) and MetricF (b) , the black dot in the circle represents user 9.

It can be seen from Fig. 6a that the independent and identically distributed feature vectors generated by VIB-DML use t-distributed Stochastic Neighbor Embedding[25] (T-SNE) to dimension reduction, and the locations of these vectors in space changes.Compared with the feature vectors generated by MetricF in Fig. 6b, the distance between the three points A, B, and C with a rating of 5 in Fig. 6a and user 9 is significantly reduced, indicating that the locations of three points A, B, and C are more precise,in accord with the relationship between ratings and distances. The above experiments show that the feature vectors generated by VIB-DML have more accurate relative positions than the feature vectors generated by MetricF. VIB-DML removes the redundant information in latent vector by restricting mutual information of the latent vector, thus the problem that the approaches of DML are easy to lead to the phenomenon of over-fitting is solved.

## 4    CONCLUSION

Aiming at the phenomenon that Euclidean distance is commonly used in distance metric learning models in recommender systems, this paper proposes a distance metric learning model VIB-DML based on variational information bottleneck to meet conditions required for Euclidean distance. The main work of this paper includes:1) Explain the limitations of the ordinary Euclidean distance,meanwhile,in combination with Variational Information Bottleneck and propose a new metric learning model named VIB-DML. 2) Make use of the characteristic that Variational Information Bottleneck can eliminate the redundancy the information, so that it can generate more robust latent vectors in the recommended scenario, and reduce the risk of model overfitting. 3) VIB-DML is compared and analyzed in many aspects through the four types of experiment. And the experimental results show that VIB-DML has two major characteristics of high accuracy and high robustness.

This paper explores how to accurately measure the relationship between users and items in the recommendation system, and improves Metric Learning using Euclidean distance.The next step is to try to use other distance functions to replace Euclidean distance,and introduce auxiliary information to transform the distance function to further improve the performance.

**Acknowledgment.** This paper is supported by the National Natural Science Foundation of China (No. 61862046),the Inner Mongolia Autonomous Region Science and Technology Achievements Transformation Project (No. CGZH2018124).